\documentclass[aps,twocolumn,superscriptaddress]{revtex4}
\usepackage{graphicx,bm,amssymb,amsmath}

\begin{document}
\title{Fingerprints of spatial charge transfer in Quantum Cascade Lasers}
\author{R. Nelander}
\email[Email:]{rikard.nelander@fysik.lu.se}
\author{A. Wacker}
\affiliation{Division of Mathematical Physics, Physics Department, Lund University, Box 118, 22100 Lund, Sweden}
\author{M. F. Pereira Jr.}
\affiliation{Materials and Engineering Research Institute, Sheffield Hallam University, Howard Street, Sheffield, S1 1WB, United Kingdom}
\author{D. G. Revin}
\author{M. R. Soulby}
\author{L. R. Wilson}
\author{J. W. Cockburn} 
\affiliation{Department of Physics and Astronomy, University of Sheffield, Sheffield S3 7RH, United Kingdom}
\author{A. B. Krysa}
\author{J. S. Roberts}
\author{R. J. Airey}
\affiliation{EPSRC National Centre for III-V Technologies, Department of Electrical and Electronic Engineering, University of Sheffield, Sheffield S1 3JD, United Kingdom}

\begin{abstract}
We show that mid infrared transmission spectroscopy of a quantum cascade laser provides clear cut information on changes in charge location at different bias. Theoretical simulations of the evolution of the gain/absorption spectrum for the $\lambda \sim$ 7.4 $\mu$m InGaAs/AlInAs/InP quantum cascade laser have been compared with the experimental findings. Transfer of electrons between the ground states in the active region and the states in the injector goes in hand with a decrease of discrete intersubband absorption peaks and an increase of broad high-energy absorption towards the continuum delocalised states above the barriers.
\end{abstract}

\maketitle

\section{Introduction}
The performance of Quantum Cascade Lasers (QCL) has improved fast since the first experimental realisation more than a decade ago \cite{FaistScience1994}. The emission wavelength now ranges between 2.95 and 190~$\mu$m (6.5 - 420 meV) \cite{DevensonAPL2007,WaltherAPL2006} and high output powers have been reached \cite{EvansAPL2004,WilliamsOE2005}. The key operating mechanism of most QCLs is efficient and rapid transport of electrons from the lower laser state to the upper laser state of the next period. In order to achieve this, the laser structure is carefully designed to reach resonance conditions at operating bias in order to optimise tunnelling rates. Knowledge of the actual location of electron charges in the structure which influences the self-consistent electric potential is crucial to design the resonance conditions as well as transition energies.

Most detailed previous theoretical studies were focused on level occupations and transport properties \cite{HarrisonAPL1999, IottiPRL2001, BonnoJAP2005, JirauschekJAP2007}. In contrast, here we focus on the evolution of the spatial distribution of charge as a function of current in combination with the optical spectra. In particular we introduce the concept of spatially resolved gain and absorption in order to better visualise the movement of charge.

Transmission spectroscopy measurements have recently been preformed to characterise the state of an operating QCL \cite{RevinAPL2006} and in this manuscript we present simulations for the same device. Preliminary results  focused on discrete transitions within the active region \cite{PereiraJMS2007}. Here we provide a more thourough discussion and additionally address absorption to the continuum and spatially resolved gain. Furthermore, we discuss some discrepancies between experiments and simulations.

This manuscript is organised as follows; section II presents our mathematical formalism. Our numerical results are presented in section III. We begin the discussion by analysing the gain spectrum (sect. IIIa) and introduce spatially resolved gain in section IIIb.  We then study high energy absorption (sect. IIIc) and discuss the discrepancies between experiments and simulation in section IIId. A short summary and conclusion close the paper (sect. IV).

\section{Theory}
Our device simulations are based on the non-equilibrium Greens functions approach (NEGF) \cite{LeePRB2002} taking into account the full non-diagonal self-energies as outlined in Ref.~\cite{LeePRB2006}. The device is described by a periodic repetition of one section of the structure, which we refer to as period. We employ a set of basis states $\Psi_{\alpha}(z)e^{i\mathbf{k}\cdot {\bf r}}$, where $\mathbf{k}$ and ${\bf r}$ are vectors in the $(x,y)$ plane perpendicular to the growth direction $z$ to describe each period. For the self-consistent transport calculations we use Wannier states \cite{WannierPR1937} for $\Psi_{\alpha}(z)$, taking into account the 14 lowest states per period together with the corresponding states of both adjacent periods. For the stationary state we calculate the eigenstates of the  free particle Hamiltonian including the mean field contribution, the Wannier-Stark (WS) states. Here we add further states (assumed to be unoccupied) in order to obtain a better resolution of the  high-frequency absorption. The quasi-energies $E_{\alpha}$ and widths $\Gamma_{\alpha}$ of the WS states are extracted from our self-consistent NEGF simulations, where the levelshifts and widths of the extra states are set in accordance to the other high-energy states.

The material gain spectra $g(\omega)$ are evaluated by a simple approach assuming Lorentzian lifetime broadening and restricting to diagonal dephasing, see also appendix~\ref{AppGain}:
\begin{multline} \label{gE}
g(\omega) = \sum_{\alpha,\beta} \frac{e^2 |z_{\alpha,\beta}|^2 
(E_\alpha - E_\beta)(n_\alpha - n_\beta)}
{2 N d \hbar c \epsilon_0 \sqrt{ \epsilon_r}} \\
          \times \frac{ \Gamma_{\alpha,\beta}}
{(E_\alpha - E_\beta -\hbar \omega)^2 + (\Gamma_{\alpha,\beta})^2/4}
\end{multline}
where $d$ is the length of one period, $e<0$ is the electron charge, $\epsilon_0$ is the dielectric constant of vacuum, $\epsilon_r$ the relative dielectric constant of the well material, $c$ is the vacuum speed of light, $E_\alpha$, $n_\alpha$ are the energy and sheet electron density of state $\alpha$, respectively, $N$ is the total number of periods considered and $\hbar \omega$ is the photon energy. Correlations in the scattering environment can reduce the absorption/gain linewidth \cite{BanitAPL2005}. However, this effect is neglected here for simplicity. In other words, the width of the individual transitions is estimated by $\Gamma_{\alpha,\beta}=\Gamma_{\alpha}+\Gamma_{\beta}$.

\section{Numerical Results and Discussion}
\begin{figure}
\includegraphics[width=1.\columnwidth]{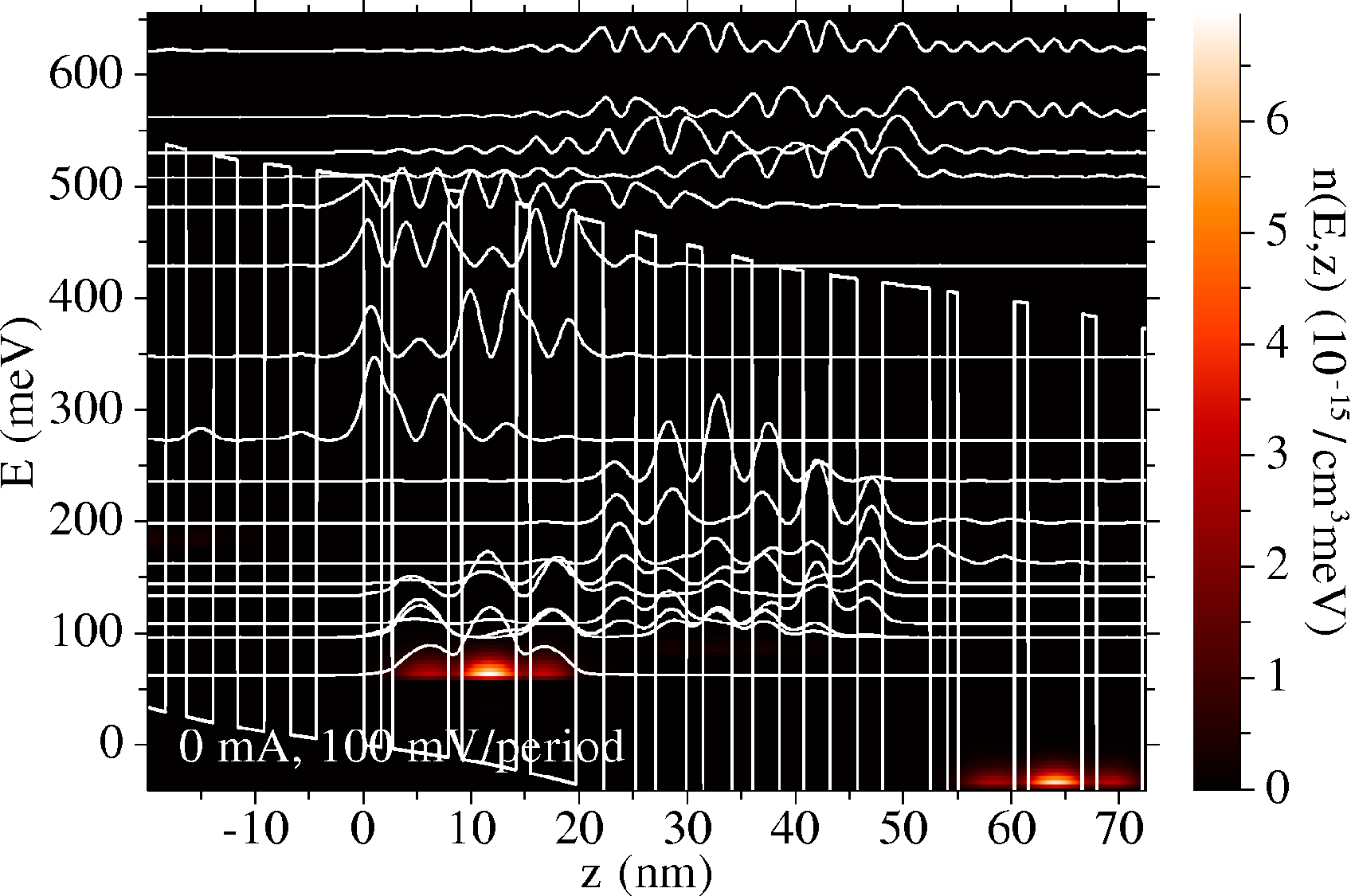}
\includegraphics[width=1.\columnwidth]{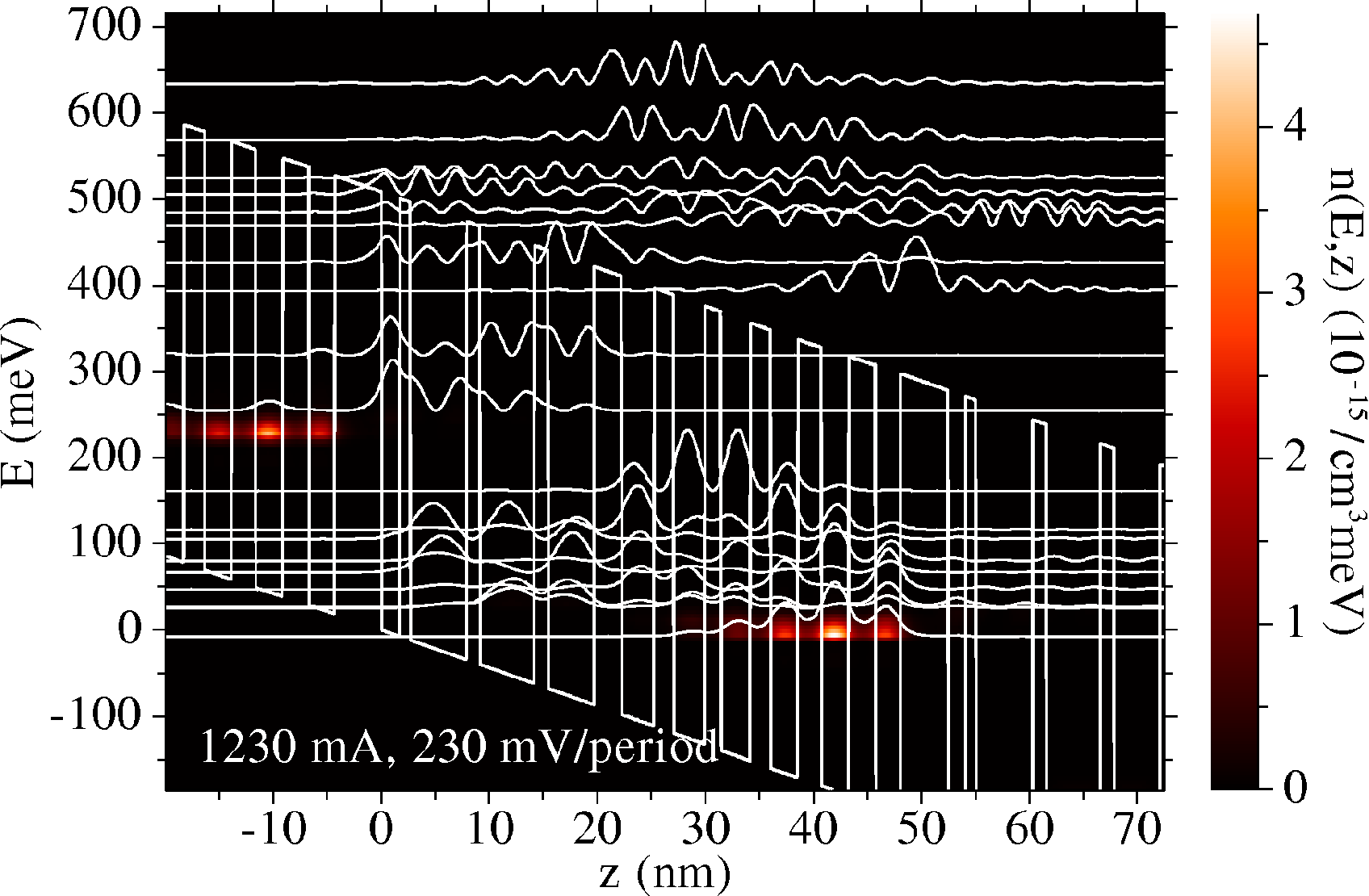}
\caption{\label{conc} Calculated energetically resolved electron concentration and modulus square of 19 Wannier-Stark states for the central period. The upper and lower panels correspond, respectively, to low and high biases.}
\end{figure}

We consider the device of Ref.~\cite{RevinAPL2006}, which is a lattice matched In$_{0.53}$Ga$_{0.47}$As/Al$_{0.48}$In$_{0.52}$As InP-based design with a four well active region emitting at $\lambda \approx 7.4$~$\mu$m (168~meV). The laser ridge is $L = 2.5$~mm long, 55~$\mu$m wide, and consists of $N=35$ periods. The nominal doping density is $9.1 \times 10^{10}$~cm$^{-2}$ per period. All calculations are performed for a lattice temperature of 77~K and we use the material parameters of Ref.~\cite{VurgaftmanJAP2001}. The conduction band profile and 19~Wannier-Stark states of the central period for two different biases can be seen in Fig.~\ref{conc}. Here we also show the energetically resolved electron distribution, which reads
\begin{equation}
n(E,z) = \frac{2(\mbox{for spin})}{2\pi A } 
\sum_{\alpha, \beta, \mathbf{k}} \psi_\beta^*(z)
\psi_\alpha(z)(-i) G^<_{\alpha, \beta}(\mathbf{k},E)\, ,
\end{equation}
where  $G^<(E)$ is the correlation function, which is related to the density matrix by $\langle \hat{a}^\dagger_\beta (\mathbf{k}) \hat{a}_\alpha (\mathbf{k}) \rangle = \tfrac{-i}{2\pi} \int \mathrm{d} E \, G^<_{\alpha, \beta}(\mathbf{k},E)$, where $\hat{a}^\dagger_\alpha (\mathbf{k})$ ($\hat{a}_\alpha(\mathbf{k})$) are the creation (annihilation) operators \cite{HaugJauhoBook1996} and $A$ is the sample area. The charge is located in the injector region around $z=10$~nm for low bias (100~mV per period), but it is transfered to the injector region around $z=40$~nm under operating conditions (230~mV per period).

\begin{figure}
\includegraphics[angle=0,width=1.\columnwidth]{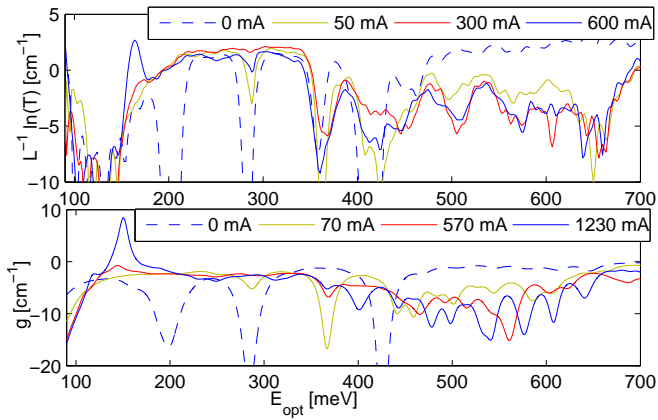}
\caption{\label{spectra} Top: Experimental modal gain estimated by transmission data as $g = \frac{1}{L} \ln T$ for TM polarisation. Gradually increased losses for the energies below about 160 meV indicate the long wavelength transmission edge for the laser waveguide. Bottom: Calculated material gain $g$.  In experiments the temperature was 10 K for zero and low current and approximately 80 K for the highest current.}
\end{figure}

The calculated and measured spectra are displayed in Fig.~\ref{spectra}. The major features are four absorption peaks at approximately 200, 280, 370, and 420~meV as well as a gain peak at 150~meV at high currents. The lower energy absorption peaks disappear with increasing current, while the broad high-energy absorption above 450 meV increases with current. As the modal gain $g_m$ is related to the material gain $g$ via  $g_m=\Gamma g -\alpha_{\rm tot}$, where $\Gamma$ is the confinement factor and $\alpha_{\rm tot}$  the total losses averaged over the whole device \cite{AgrawalBook1986}, the experimental data and the calculated results should have the same spectral shape, but do not need to agree fully quantitatively.

\begin{figure}
\includegraphics[angle=0,width=1.\columnwidth]{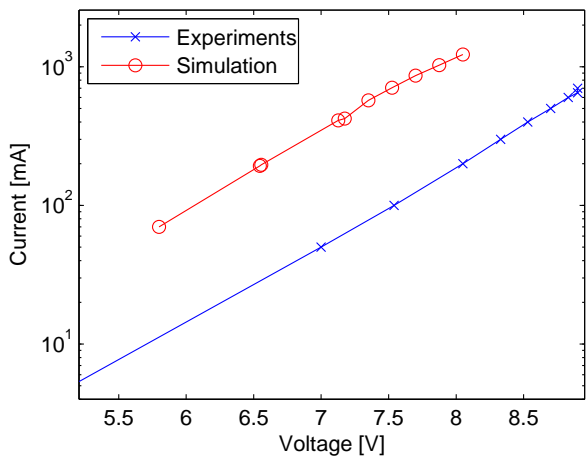}
\caption{\label{IV} Calculated and measured current-voltage characteristic. While the voltage in the simulation is defined as the voltage drop per period times the number of periods, the experimental values include the voltage drop over the contact regions as well. For small biases the calculated currents are not sufficiently precise for a reasonable display on a logarithmic scale.}
\end{figure}

In Fig.~\ref{IV} the calculated and measured current-voltage relation is shown. Although the slope on a logarithmic current-scale is similar there is still a substantial difference between experiments and theory; almost one order of magnitude in current or 20~\% in voltage.

\subsection{Gain Spectrum}
The reason for the bias dependence of the absorption peaks (see Fig.~\ref{spectra}) can be understood by an analysis of Fig.~\ref{conc}: For low bias, (top panel of Fig.~\ref{conc}), the carriers essentially occupy the ground state in the active region $0 \lesssim z \lesssim 20$~nm, and the absorption peaks at 200, 280, 370, and 420~meV, can be related to the transitions to the excited states in the active region. However, the absorption peak at 370~meV is absent at zero current in our simulation. Here we have an almost vanishing dipole matrix element due to the approximate even parity of both the occupied state and the state at 430~meV in the upper Fig.~\ref{conc}. As this peak is present in the experiment, the real structure must differ somewhat. In addition there is some high energy-absorption by transition towards the high energy states above the band edges of the barriers. However the matrix elements between the well-localised ground state and these fairly delocalised states are small.

For high bias and current, bottom of Fig.~\ref{conc}, the charge is transferred to the injector region. Thus the main absorption peaks at 200, and 280 meV are bleached. The higher energy absorption peaks at 370, and 420~meV exhibit a non-monotonic behaviour with current, as the corresponding excited levels mix with continuum states.

The movement of charges to the injector leads to a significant increase of high energy absorption, since the states in the injector region are more  extended and thus yield larger dipole moments for transitions to the continuum states.

\subsection{Spatial resolution of gain}
The location of charge density as a function of bias is explicitly visualised by the spatial resolved gain plotted in Fig.~\ref{gainEZ}, which is calculated with the following expression (see appendix~\ref{AppGain}),
\begin{eqnarray} \label{gEZ}
g_m(\omega, z) &=&  \frac{\hbar e^2}{2 c \epsilon_0 \sqrt{ \epsilon_r} } 
\sum_{\alpha \beta}  \frac{\psi^*_\alpha(z) \partial_z \psi_\beta(z) -  
\partial_z  \psi^*_\alpha(z) \psi_\beta(z)}{2 m_e(z)} \nonumber \\
&\times& z_{\beta \alpha} (n_\alpha - n_\beta) 
\frac{\Gamma_{\beta,\alpha}}{(E_{\alpha} - E_{\beta} + \hbar \omega)^2 + \Gamma^2_{\beta,\alpha}/4}.
\end{eqnarray}
Here $\partial_z \psi(z)$ the derivative of $\psi(z)$ with respect to $z$.

One can clearly see that most absorption/gain within the energy range $100 \, \, \mathrm{meV} < E_{\rm opt} < 350 \, \, \mathrm{meV}$ is located in the active region while the higher energy absorption follows the electron density position, i.e. localised in the active region for small currents and to the injector region for high currents. A single transition oscillates spatially between gain and absorption. However, integrating over $z$ provides the total absorption from Fig.~\ref{spectra}. The gain transition at 150~meV in the lower part Fig.~\ref{gainEZ} only shows gain contributions. This demonstrates the successful optimisation of the dipole matrix element of the lasing transition during the design of the laser structure.

\begin{figure}
\includegraphics[width=1.\columnwidth]{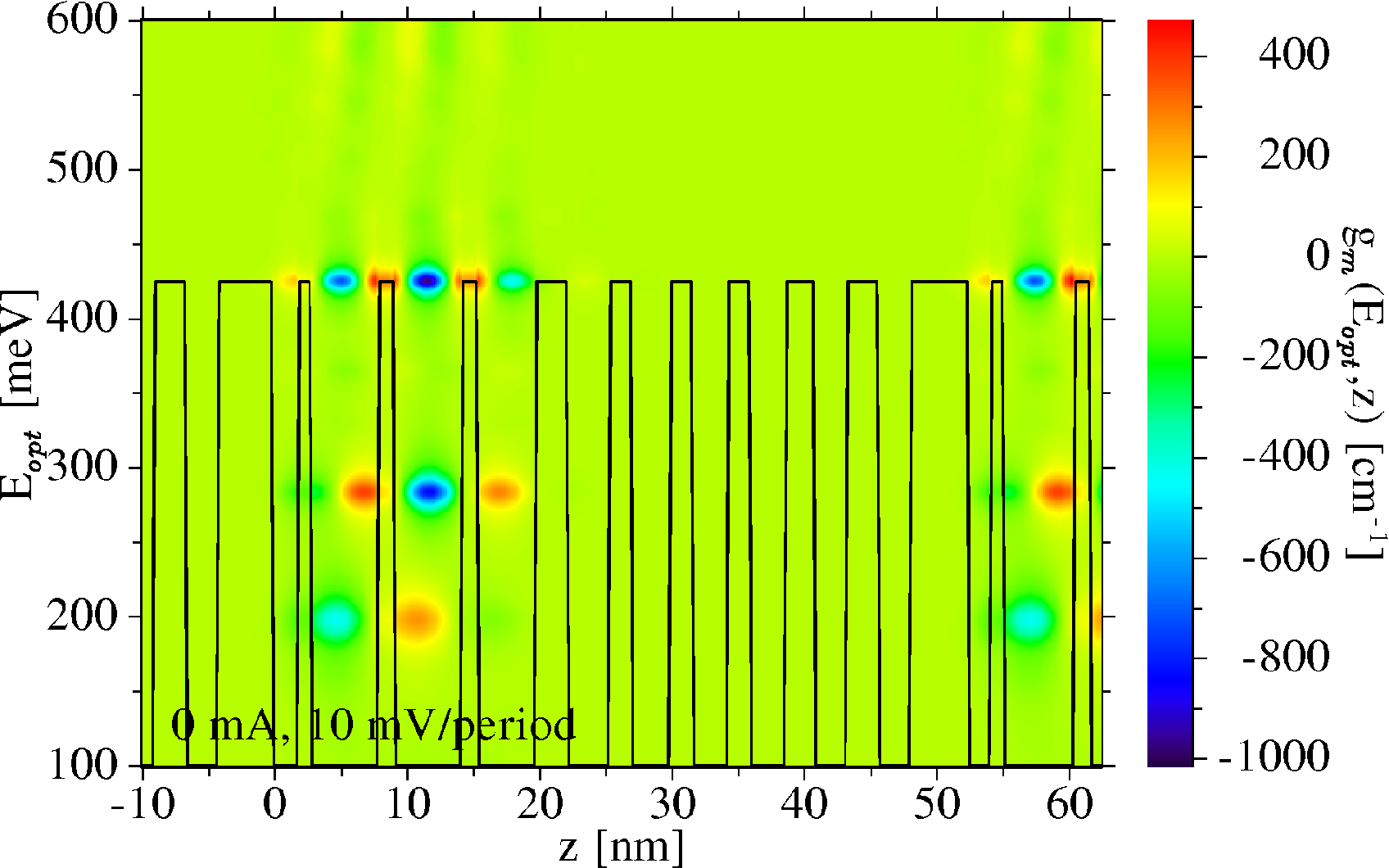}
\includegraphics[width=1.\columnwidth]{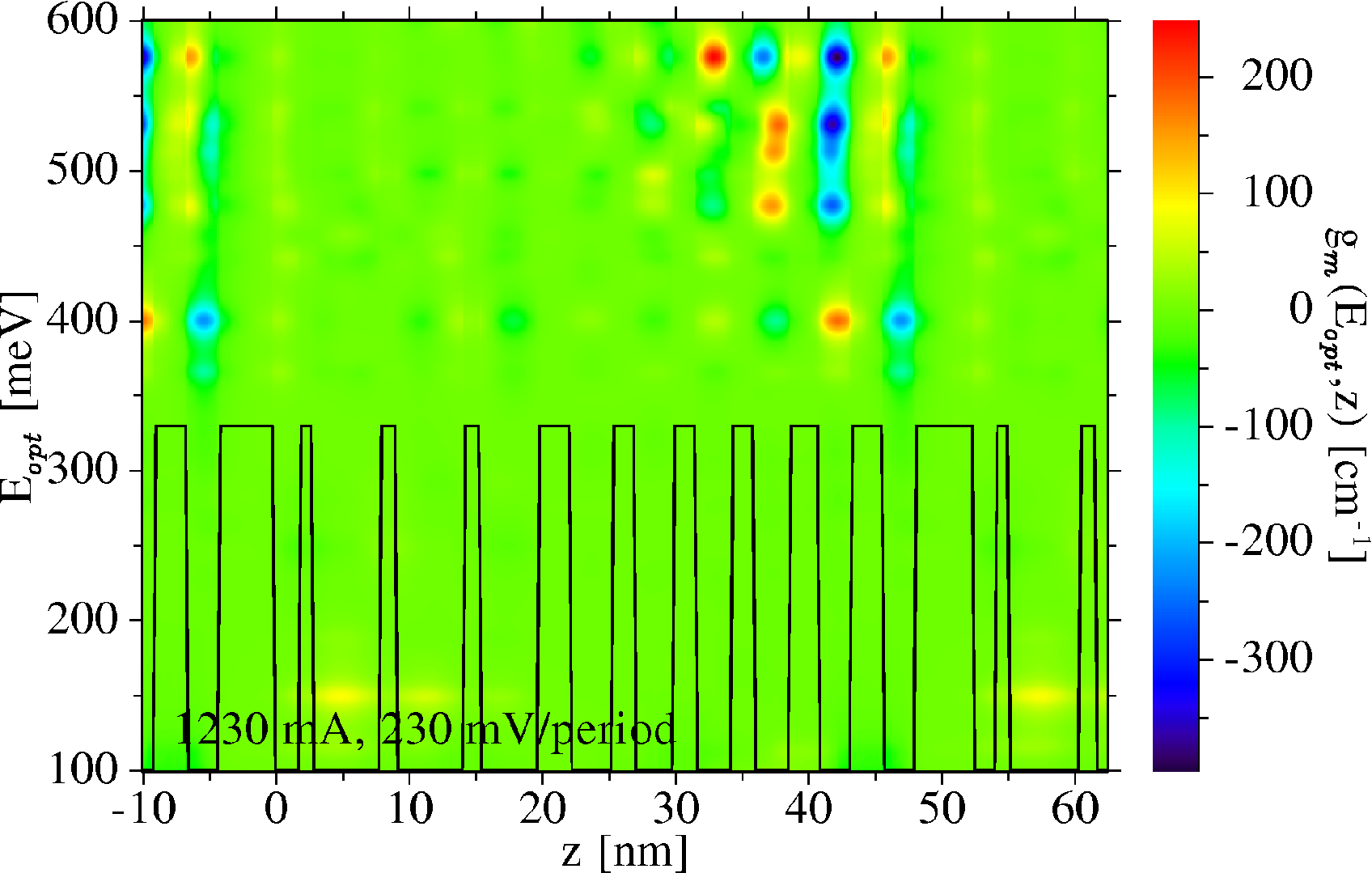}
\caption{\label{gainEZ} Spatial resolved gain for a low bias (top panel) and a high bias (bottom panel). Below 100~meV there is extremely strong absorption (on this scale) in the injector region. The conduction band profile in the plot is merely for spatial orientation, while its energy is arbitrary.}
\end{figure}

\subsection{High energy absorption}
As discussed above, the absorption for photon energies above $E_{\rm opt}=400$~meV is a marker for the location of the electron charge, as the injector states are more extended compared to the ground state in the active region. Accordingly the dipole transition matrix elements with the states from the continuum above the barriers are larger. While this trend is clear both in experiment and simulation, the detailed structure differs.

\begin{figure}
\includegraphics[angle=0,width=1.\columnwidth]{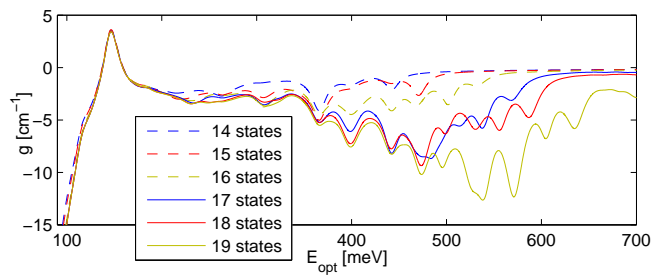}
\caption{\label{specN} Calculated spectrum for 1030 mA with different number of Wannier States. The gain peak height is correctly described by the 14 states included in the transport simulation while the high energy absorption changed drastically when adding more high energy states.}
\end{figure}

We believe that this is due to two reasons. First, the calculated absorption in this range strongly depends on the number of states taken into account. This is shown in Fig.~\ref{specN}, where one can see, that the spectra converges with increasing number of states only for energies up to about 500 meV. Second, the quantitative determination of these high energy states is less straightforward, as standard models such as effective mass or $\mathbf{k \cdot p}$ models, are less accurate at these high energies. Other minima than the $\Gamma$-point might also come into play.

To our knowledge, high energy absorption in QCLs has not been studied in detail. With the common approach to calculate the Wannier-Stark states directly for a QCL embedded in a finite box, the high energy states are difficult to access, as their extended nature is not compatible with an artificial boundary. The difficulty is overcome by our approach starting with Wannier states \cite{LeePRB2002}, where the spatial cutoff is replaced by a cutoff in energy with respect to the local conduction band. While transitions to these states can shed light on the localisation of charges, we did not find any indication, that they are of relevance for transport properties or the gain spectrum around 150 meV. Thus they can be safely neglected as long as only these standard properties are of interest.

\subsection{Discrepancies in current}

Comparing simulations and experiments we observe good overall agreement. However there is one main discrepancy: the calculated current is substantially larger than the experimental value. We attribute these results partly to the fact that the parameters used in calculations slightly differ from the real sample parameters and partly to the limitations of the theoretical method used. The voltage drop over the device region and the effective electron concentration are the main less-known experimental parameters.

The voltage drop $V_{\rm contact}$ outside the QCL structure, such as contact regions and cladding layers may add up to about one 
volt or more, which could explain the difference in the current-voltage characteristics, Fig.~\ref{IV}, where $V_{\rm exp}=NFd+V_{\rm  contact}$, while  $V_{\rm sim}=NFd$, where $F$ is the average electric field over the device region and $Nd = 1.83$~$\mu$m is the total length of the QCL structure. Also there is some uncertainty regarding the real doping level that could be slightly higher or lower than the nominal value.

Judging from Fig.~\ref{IV} and assuming that many-body effects are small (like electron-electron interaction and the Pauli principle) the relation 
\begin{equation}
I \propto n \, e^{F/F_0}
\label{EqEstI}
\end{equation}
seems reasonable, where $I$ is the current, $n$ the electron concentration, and $F_0$ is a constant. This suggest that a small error in the actual field can give a large error in the current, while an error in the electron concentration gives the same relative error in the current. We can therefore conclude that the discrepancy in current-voltage between experiments and simulation can be explained both by a small voltage drop over the contacts and cladding region and an uncertainty in the actual doping density.

The experimental gain $g_m(\omega)$ in Fig.~\ref{spectra} is calculated from transmission data $T(\omega)$ with $g(\omega) = \frac{1}{L} \ln T(\omega)$, where $L$ is the length of the laser ridge, 2.5~mm in this device. This value differs from the calculated material gain $g$ due to the confinement factor and the frequency-dependent losses, which makes a quantitative comparison difficult.

The same parameter uncertainty as in the current-voltage case affect to the overall magnitude of gain and absorption peaks, which are somewhat larger in the simulations. If we assume, as previously, that many-body effects are small, a simple relation 
\begin{equation} \label{EqEstg}
g(\omega) \propto n \, f(F,\omega)
\end{equation}
holds. Thus the onset of gain is related to a fixed field $F_{\rm onset}$ and for currents, where the gain spectrum exhibits qualitative changes, quantitative information can be extracted. In particular, the gain peak at 150 meV sets in at $I_{\rm exp}\approx 600$~mA, while this feature appears in the simulations for $I_{\rm sim} \approx 1000$~mA, which suggest a smaller effective electron density. Another aspect refers to the width of the gain/absorption peaks, which are larger in the simulation. This indicates the presence of correlations in the scattering environment for the states involved \cite{BanitAPL2005}, which we neglected in our simple model.

A last remark concerns the position of the gain peak. The positions of the absorption peaks in simulation agree excellent with experiments, while in experiments the gain peak is 15~meV higher in energy than in simulation. This may be related to pronounced low frequency absorption features (such as free carrier absorption in the cladding regions) not taken into account in the simulation. Such a mechanism reduces the left side of the gain peak stronger than the right side thus shifting the observed gain peak position to higher frequencies.

\section{Conclusion}
The evolution of the gain/absorption spectrum of a quantum cascade laser provides a clear signal for the spatial transfer of charge inside the structure. If the electrons dominantly occupy a localised level (such as the ground state in the active region), one observes several distinct peaks due to transitions to higher levels in the same region, but only weak absorption to the delocalise states above the barriers. This continuum absorption becomes more pronounced for transitions from the more extended injector states, while the aforementioned absorption peaks vanish, if the majority of charge is transfered hither.

These features can be described by resolving the gain spatially, Eq.~(\ref{gEZ}). Our simulations are consistent with the experimental findings if we assume some uncertainty in the actual electron density in the sample and that a part of the voltage drops outside the actual QCL structure.

\section{Acknowledgement}
This work was supported by the Swedish Research Council (VR) and the European Union Marie Curie Research Training Network "POISE".

\appendix

\section{Spatially resolved gain}\label{AppGain}
In order to pin-point the spatial location of different spectroscopic features without resorting to investigating every possible transition an expression for a spatially resolved gain is sought. Our starting point is the common relation \cite{Jackson1998}
\begin{eqnarray}
g(\omega,z) \approx - \frac{\Re \{ \sigma(\omega,z) \} }{c \epsilon_0 \sqrt{\epsilon_r }}
\end{eqnarray}
where we have taken into account a $z$-dependence of  the gain $g(\omega,z)$ and conductivity $\sigma(\omega,z)$. The expression have a simple intuitive explanation: if the electric field and current are in phase, energy is accumulated in the sample corresponding to absorption (also known as Joule heating), and vice versa.

The TM mode exhibits a small oscillating electric field, $\delta F(\omega)$, which is homogeneous on the length scale of a period of the structure.  This implies the change \begin{widetext}
\begin{equation}
\sum_\mathbf{k} \int \frac{\mathrm{d} E}{2 \pi} \delta G^<_{\alpha \beta,  \mathbf{k}}(E) = \frac{Ae}{2} z_{\beta \alpha} (n_\alpha - n_\beta) \delta F(\omega) \frac{1}{E_{\alpha} - E_{\beta} + \hbar \omega + i \Gamma_{\beta,\alpha}/2} 
\end{equation}
for the Green's functions calculated in linear response from Ref.~\cite{BanitAPL2005} neglecting the $\delta \Sigma$ terms. The local current, $\delta J(\omega,z)$ is then given by \cite{LeePRB2006}
\begin{equation}
\delta J_0(E,z) = \frac{-e}{2 \pi A} \sum_{\alpha \beta, \mathbf{k}} \left(  \frac{\hbar}{m^*(z)} \psi^*_\alpha(z) \frac{\partial   \psi_\beta(z)}{\partial z} - \frac{\hbar}{m^*(z)} \frac{\partial \psi^*_\alpha(z)}{\partial z} \psi_\beta(z) \right)  \delta G^<_{\alpha \beta, \mathbf{k}}(E)  \, .
\end{equation}
\end{widetext}
Integrating over $E$ and using $\sigma(\omega,z) = \delta J_0(\omega,z)/ \delta F(\omega)$ we obtain Eq.~(\ref{gEZ}). Averaging Eq.~(\ref{gEZ}) over $z$ we obtain Eq.~(\ref{gE}) by using
\begin{equation}
\frac{1}{Nd} \int \mathrm{d} z \, \psi^\dagger_\alpha(z) \frac{- i \hbar}{m^*(z)} \partial_z \psi_\beta(z)
 = \frac{1}{Nd} \frac{i}{\hbar} \left[ H,z\right]_{\alpha,\beta} 
\end{equation}

\end{document}